\documentclass[twocolumn,showpacs,prd]{revtex4}
\usepackage{mathrsfs,bm}
\usepackage{longtable,lscape}
\usepackage{txfonts}
\usepackage{amssymb}
\usepackage{indentfirst}
\usepackage{graphicx,,booktabs}
\usepackage{color}
\usepackage{amssymb}

\begin{document}

\title{The open-charm radiative and pionic decays of molecular charmonium $Y(4274)$}
\author{Jun He$^{1,2}$}\email{junhe@impcas.ac.cn}
\author{Xiang Liu$^{1,3}$\footnote{Corresponding author}}\email{xiangliu@lzu.edu.cn}

\affiliation{
$^1$Research Center for Hadron and CSR Physics,
Lanzhou University and Institute of Modern Physics of CAS, Lanzhou 730000, China\\
$^2$Nuclear Theory Group, Institute of Modern Physics of CAS, Lanzhou 730000, China\\
$^3$School of Physical Science and Technology, Lanzhou University, Lanzhou 730000,  China}
\date{\today}

\begin{abstract}
In this work, we investigate the decay widths and the line shapes of the open-charm radiative and pionic decays of $Y(4274)$ with
the $D_s\bar{D}_{s0}(2317)$ molecular charmonium assignment. Our calculation indicates that the decay widths of $Y(4274)\to D^{+}_{s}D^{*-}_{s}\gamma$ and $Y(4274)\to D^+_{s}D^-_{s}\pi^0$ can reach up to 0.05 keV and 0.75 keV, respectively. In addition, the result of the line shape of the photon spectrum of $Y(4274)\to  D_s^+ {D}_s^{*-} \gamma$ shows that there exists a very sharp peak near the large end point of photon energy. The line shape of the pion spectrum of $Y(4274)\to D_s^+ {D}_s^{*-} \pi^0$ is similar to that of the pion spectrum of $Y(4274)\to  D_s^+ {D}_s^{*-} \gamma$, where we also find a very sharp peak near the large end point of pion energy. According to our calculation, we suggest further experiments to carry out the search for the open-charm radiative and pionic decays of $Y(4274)$.
\end{abstract}

 \maketitle
\maketitle

\section{introduction}\label{sec1}

Recently the CDF Collaboration reported an explicit enhancement
structure with 3.1$\sigma$ significance in the $J/\psi\phi$ invariant
mass spectrum of the $B^+\to K^+ J/\psi\phi$ process. In this work, we refer to this new enhancement structure by the name $Y(4274)$. Its mass and width are  $M=4274.4^{+8.4}_{-6.7}(\mathrm{stat})\pm1.9(\mathrm{syst})$ MeV/c$^2$ and $\Gamma=32.3^{+21.9}_{-15.3}(\mathrm{stat})\pm7.6(\mathrm{syst})$ MeV/c$^2$ \cite{Collaboration:2011at}, respectively. We need to specify that this new result appearing in the $J/\psi\phi$ invariant mass spectrum is based on a sample of $p\bar{p}$ collision data at $\sqrt{s}=1.96$ TeV with an integrated luminosity of about $6.0$ fb$^{-1}$~\cite{Collaboration:2011at}.
Additionally, the experiment presented in Ref. \cite{Collaboration:2011at} also confirmed the observed $Y(4140)$ previously announced in Ref. \cite{Aaltonen:2009tz}.

Before finding the $Y(4274)$ structure, there had been six charmonium-like states observed in $B$ meson decays, which include $X(3872)$ in $B\to \underline{J/\psi\pi^+\pi^-} K$~\cite{Choi:2003ue}, $Y(3940)$ in $B\to \underline{J/\psi\omega} K$~\cite{Abe:2004zs,Aubert:2007vj}, $Y(4140)$ in $B\to \underline{J/\psi\phi} K$~\cite{Aaltonen:2009tz}, $Z^+(4430)$ in $B\to \underline{\psi^\prime \pi^+} K$~\cite{:2007wga}, and $Z^+(4015)$ and $Z^+(4248)$ in $B\to \underline{\chi_{c1} \pi^+} K$ \cite{Mizuk:2008me}, where we use the underlines to mark the corresponding decay channels of charmonium-like states observed in experiments. The evidence of $Y(4274)$ revealed by CDF \cite{Collaboration:2011at} not only has made the spectroscopy of charmonium-like states observed in $B$ meson decays abundant, but has also stimulated theorists' interest in revealing its underlying structure. Studying $Y(4274)$ will improve our understanding of the essential mechanism resulting in these structures. Very recently, LHCb also confirmed the observation of $Y(4274)$ with $53\pm$ events in the $B^+\to J/\psi\phi K^+$ decay \cite{Aaij:2012pz}.

Since $Y(4274)$ was observed in the $J/\psi\phi$ invariant mass spectrum, we can conclude that the quantum numbers of $Y(4274)$ are $I^G(J^{PC})=0^+(J^{++})$ with $J=0,1,2$ if $Y(4274)\to J/\psi \phi$ occurs via S-wave.  If explaining $Y(4274)$ as a candidate of charmonium, $Y(4274)$ should be a P-wave state with the second radial excitation. In Ref. \cite{Liu:2009fe}, the predicted total widths of the second radial excitations of $\chi_{c0}$ and $\chi_{c1}$ are larger than the width of $Y(4274)$. In addition,
$X(4350)$, which was observed in the $\gamma\gamma\to \phi J/\psi$ process with mass $m=4350.6^{+4.6}_{-5.1}(\mathrm{stat})\pm0.7(\mathrm{syst})$ MeV and width $\Gamma=13^{+13}_{-9}(\mathrm{stat})\pm 4(\mathrm{syst})$ MeV \cite{Shen:2009vs}, was explained as the candidate of the second radial excitation of $\chi_{c2}$ \cite{Liu:2009fe}, where
the measured parameters of $Y(4274)$ are different from those of $X(4350)$.
However, at present we cannot fully exclude the P-wave charmonium explanation of $Y(4274)$, since the uncertainty of the quark pair creation model is not under control \cite{Liu:2009fe}.

Besides the conventional charmonium assignment to $Y(4274)$, $Y(4274)$ as molecular charmonium can be produced by $B$ meson decay. The comparison of the existing experimental information of charmonium-like states observed in $B$ meson decays reflects a common property; i.e., these charmonium-like states are near the threshold of the corresponding charmed meson pair, which has provoked the investigation of whether these observed charmonium-like states $X(3872)$, $Z^+(4430)$, $Z^+(4015)/Z^{+}(4248)$, $Y(3930)$, $Y(4140)$, and $Y(4274)$ can be explained as the corresponding molecular charmonia \cite{Swanson:2003tb,Liu:2008fh,Liu:2008tn,
Liu:2007bf,Liu:2008xz,Shen:2010ky,Ding:2008gr,Ding:2009zq,Liu:2009ei,Liu:2010hf,Liu:2009pu}.
The production of molecular charmonium via the $B$ meson decay is allowed. The $c\bar{c}$ pair is created from the color-octet mechanism in the weak decays of the $B$ meson; then $c$ and $\bar{c}$, respectively, capture $\bar{q}$ and $q$ to form a charmed meson pair, where a color-octet $q\bar{q}$ pair is popped out by a gluon. Thus, a pair of the charmed mesons with low momentum easily interact with each other to form the molecular charmonium \cite{Liu:2010hf}.

As indicated in Refs. \cite{Shen:2010ky,Liu:2010hf}, the mass of $Y(4274)$ is
near the threshold of $D_s\bar{D}_{s0}(2317)$, which is similar to the situations
of $Y(4140)$ and $Y(3930)$, since $Y(4140)$ and $Y(3930)$ are assigned as molecular states of
$D^*_{s}\bar{D}^*_{s}$ and $D^*\bar{D}^*$, respectively \cite{Liu:2010hf}. Thus, it is natural to deduce that $Y(4274)$ can be an S-wave $D_s\bar{D}_{s0}(2317)$ molecular charmonium with the flavor wave
function
\begin{eqnarray}
|Y(4274)\rangle=\frac{1}{\sqrt{2}}\big[|D_s^{+}D_{s0}^{-}\rangle
+|D_s^{-}D_{s0}^{+}\rangle\big],\label{wavefunction}
\end{eqnarray}
which is also supported by the dynamical calculation of its mass when assuming $Y(4274)$ as an S-wave
$D_s\bar{D}_{s0}(2317)$ molecular state with spin-parity $J^P=0^-$ \cite{Shen:2010ky,Liu:2010hf}.

Thus, studying $Y(4274)$ under different assignments will be helpful in distinguishing its possible structure explanations. In this work, we mainly focus on the decay behavior of $Y(4274)$ as a charmonium-like molecular structure. Performing the study of the decay behavior of $Y(4274)$ can provide crucial information for testing this molecular assignment to $Y(4274)$. Considering only the
preceding reasons and the present theoretical research status of $Y(4274)$, in this work we investigate the open-charm radiative and pionic decays of $Y(4274)$, which includes the calculation of the branching ratios and the study of the line shape of these decays.
Our study will give hints for further experimental studies of $Y(4274)$, especially in searching for other decay channels of $Y(4274)$.

This work is organized as follows. After this introduction, we illustrate the calculation details of the open-charm radiative and pionic decays of $Y(4274)$ under the assignment of molecular charmonium.
In Section \ref{sec3}, the numerical result will be presented. The paper ends with the discussion and conclusion.

\section{Radiative and pionic open-charm decays of $Y(4274)$}\label{sec2}

Under the molecular charmonium assignment to $Y(4274)$, it is
interesting to investigate its radiative and pionic decays, since the photon and pion, respectively, from the
radiative and pionic decays of $Y(4274)$ can be easily detected in
experiments. As a realistic research topic, both a theoretical
estimate of their branching ratios and a study of the line shapes
of the photon and pion spectra of the corresponding decays can
reflect the internal structure of
$Y(4274)$ \cite{Liu:2009pu,Voloshin:2003nt,Voloshin:2005rt}.

\begin{figure}[htb]
\begin{center}
\begin{tabular}{ccc}
\includegraphics[scale=0.45]{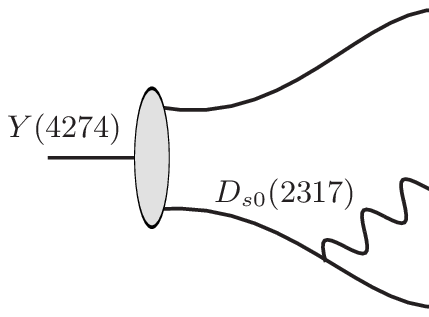}&\hspace{1cm}\raisebox{-1ex}{\includegraphics[scale=0.45]{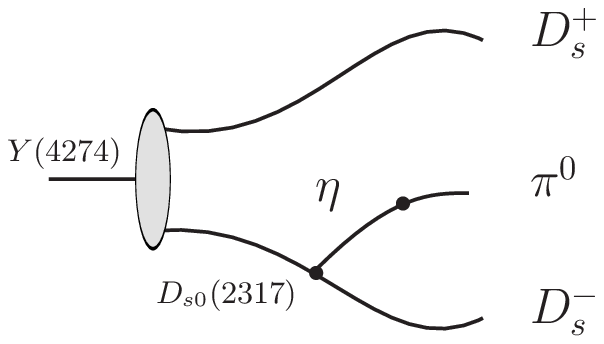}}&
\raisebox{-1ex}{\includegraphics[scale=0.45]{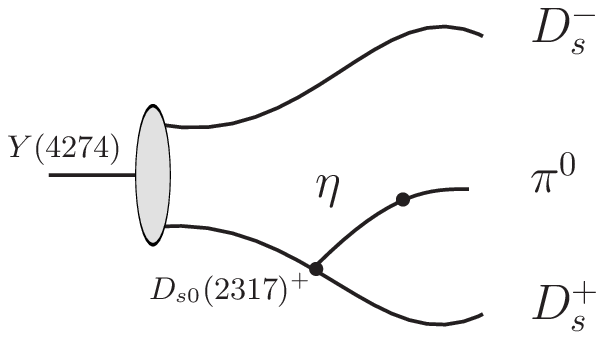}}
\end{tabular}
\end{center}
\caption{The radiative and pionic open-charm decays of
$Y(4274)$.
\label{decay1}}
\end{figure}

In this work, we focus on two groups of decay channels of $Y(4274)$, i.e., the radiative ($D_s^+D_s^{*-}\gamma$ and $D_s^-D_s^{*+}\gamma$) and pionic ($D_s^+D_s^-\pi^0$) open-charm decays. In what follows, we take the decays $Y(4274)\to
D_s^+D_s^{*-}\gamma$ and $Y(4274)\to D_s^+D_s^-\pi^0$ as examples to illustrate the relevant calculation. Under the $D_s^+{D}_{s0}(2317)^-+h.c.$ molecular state assignment, $Y(4274)$ first dissociates into $D_s+{D}_{s0}(2317)^-$ or ${D}^-_s+{D}_{s0}(2317)^+$. Then,
the radiative decay $Y(4274)\to D_s^+{D}_{s}^{*-}\gamma$ and strong decay $Y(4274)\to D_s^+\bar{D}_{s}^{-}\pi^0$ occur via the transitions ${D}_{s0}(2317)^-\to \bar{D}_{s}^{*-}\gamma$ and ${D}_{s0}(2317)^-\to{D}^-_{s}\pi^0/{D}_{s0}(2317)^+\to {D}_{s}^+\pi^0$, respectively, where ${D}_{s0}(2317)^-/{D}_{s0}(2317)^+$ decays into ${D}_{s}^-\pi^0/{D}_{s}^+\pi^0$ by the $\eta-\pi^0$ mixing mechanism \cite{Aubert:2003fg,Wei:2005ag,Liu:2006jx}. The hadron-level descriptions of $Y(4274)\to D_s^+{D}_{s}^{*-}\gamma$ and $Y(4274)\to D_s^+{D}_{s}^{-}\pi^0$ are shown in Fig.~\ref{decay1}.

The general expressions of the transition matrix elements for the radiative and pionic open-charm decays of
$Y(4274)$ can be expressed as
\begin{eqnarray}
    &&\mathcal{M}[Y(4274)\to D^{+}_{s}D^{*-}_{s}\gamma]\nonumber\\
    &&=\langle D^{*-}_{s}\gamma
    |\mathcal{H}_\gamma|{D}_{s0}(2317)^-\rangle
    \langle {D}_{s0}(2317)^-D^+_{s} |\mathcal{H}_Y|Y(4274)\rangle,\ \ \ \label{f1}
\\
    &&\mathcal{M}[Y(4274)\to D^+_{s}D^-_{s}\pi^0]\nonumber\\
    &&=\langle D_{s}^-\pi^0 |\mathcal{H}_\pi|{D}_{s0}(2317)^-\rangle
    \langle {D}_{s0}(2317)^-D^+_{s}
    |\mathcal{H}_Y|Y(4274)\rangle\nonumber\\
    &&\quad+\langle D_{s}^+\pi^0 |\mathcal{H}_\pi|{D}_{s0}(2317)^+\rangle
    \langle {D}_{s0}(2317)^+D^-_{s}
    |\mathcal{H}_Y|Y(4274)\rangle,\nonumber\\ \label{f2}
\end{eqnarray}
where $\mathcal{H}_Y$ describes the collapse of the S-wave
$D_{s}\bar{D}_{s0}(2317)$ molecular state into $D_{s}^+$ and
${D}_{s0}(2317)^-$. $\mathcal{H}_\gamma$ or $\mathcal{H}_\pi$
denotes the interaction of ${D}_{s0}(2317)^-$ with
$D^{*-}_{s}\gamma$ or $D^-_s\pi^0$. The main task of this work is to
obtain $\langle {D}_{s0}(2317)^\pm D^\mp_{s}|\mathcal{H}_1|Y\rangle$,
which describes the collapse of the molecular charmonium $Y(4274)$,
where we adopt the covariant spectator theory (CST)
\cite{Gross:1969rv,Gross:1972ye,Buck:1979ff,Gross:1982ny,Gilman:2001yh,Stadler:1996ut}
to deduce it.

\subsection{Covariant Spectator Theory}\label{T1}

The CST was proposed and developed to study the wave functions and the
form factor of deuteron
~\cite{Gross:1969rv,Gross:1972ye,Buck:1979ff,Gross:1982ny,Gilman:2001yh,Stadler:1996ut},
where the CST is also referred to as the Gross equation. To some extent, the CST is an equivalent description of the Bethe-Salpeter (BS) equation when both are solved exactly.
Thus, the CST is also widely applied to study two-body bound states. Since one particle is set to be on-shell, the Gross equation can be written easily in a form depending only on three
momenta, which is different from the case of the BS equation.

\begin{figure}[h!]
\begin{center}
\includegraphics[bb=80 580 540 720,scale=0.5]{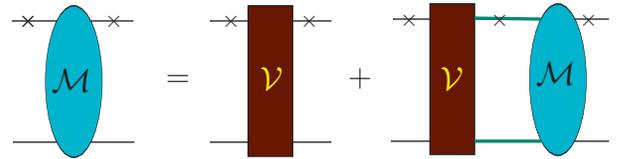}
\end{center}
\caption{(Color online.) Diagrammatic representation of the Gross equation for the two-body scattering
matrix. Here ``$\times$'' denotes the particle on-shell.
 \label{BS}}
\end{figure}

The Gross equation reads as
\begin{eqnarray}
    {\cal M}={\cal V}+{\cal V}G{\cal M}\label{h2}
\end{eqnarray}
where $\cal V$ is the two-body interaction kernel and $G$ is the propagator with particle 1 on-shell.
In the CST, the Gross equation is shown in Fig.~\ref{BS}.

\begin{figure}[h!]
\begin{center}
\includegraphics[bb=60 600 540 720,scale=0.5,clip]{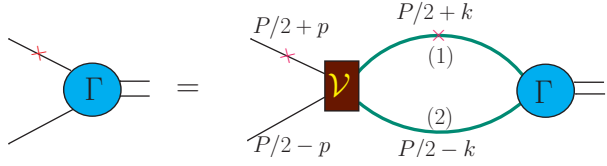}
\end{center}
\caption{(Color online.) Diagrammatic representation of the Gross equation for
the vertex function $\Gamma$. Here we use $(1)$ and (2) to mark particle 1 and particle 2, respectively.
 \label{V}}
\end{figure}

Thus, bound state equations emerge automatically from the Gross
equation. If the two-body system has a bound state at $P^2 = M_R^2$,
one has
\begin{eqnarray}
{\cal M}=\frac{|\Gamma\rangle\langle \Gamma|}{M_R^2-P^2}+{\cal R},\label{h1}
\end{eqnarray}
where $|\Gamma\rangle$ is the vertex function and $\cal R$ is finite at $P^2 =
M_R^2$. By Eqs. (\ref{h2}) and (\ref{h1}), we obtain
\begin{eqnarray}
    &&|\Gamma\rangle=VG|\Gamma\rangle,
\end{eqnarray}
which corresponds to the description in Fig.~\ref{V}.

After integrating over $k^0$, one obtains that the Gross equation only depends on
three-momentum
\begin{eqnarray}
    \Gamma({\bm p})&=&\int\frac{d^3
    k}{(2\pi)^3} {\cal V}({\bm p},{\bm k},W)
    G({\bm k},W)~\Gamma({\bm k}),
    \label{Eq: Gamma}
\end{eqnarray}
where $P=(W,\bm 0)$ is the four-momentum of a two-body system. $p$ and
$k$ are the relative momenta as shown in Fig.~\ref{V}.
Since particle 1 is on the mass shell,
${k}=({k}_0,\bm k)$,
${k}_0=E_{1}({\bm k})-\frac{1}{2}W$, and $E_1({\bm k})=\sqrt{m_1^2+{\bm k}^2}$.
${\cal V}({\bm p},{\bm k},W)$ is the interaction
kernel with particle 1 on the mass shell.

For the case of $Y(4274)$, the two-body propagator $G(\bm k,W)$ in Eq.~(\ref{Eq: Gamma}) is
\begin{eqnarray}
    G({\bm k},W)&=&
    \frac{1}{4E_{1}({\bm k})E_{2}({\bm k})}
   \bigg[\frac{1}{E_{2}({\bm k})-E_{1}({\bm k})+W}\nonumber\\&&+\frac{1}{E_{2}({\bm k})+E_{1}({\bm
    k})-W}\bigg].
\end{eqnarray}

The wave functions of the bound state can be introduced by
\begin{eqnarray}
\psi^+({\bm
p})&=&\frac{1}{\sqrt{(2\pi)^3 2W}}\nonumber\\
    &&\times
\frac{\Gamma(\bm p)}{\sqrt{2E_{1}({\bm p})2E_{2}({\bm p})} \left[E_{2}({\bm p})+E_{1}({\bm p})-W\right]},\ \ \label{Eq: wf1}\\
\psi^-({\bm
p})&=&\frac{1}{\sqrt{(2\pi)^3 2W}}\nonumber\\
    &&\times
\frac{\Gamma(\bm p)}{\sqrt{2E_{1}({\bm p})2E_{2}({\bm p})} \left[E_{2}({\bm p})-E_{1}({\bm p})+W\right]}.\ \ \label{Eq: wf}
\end{eqnarray}

By the definition of wave functions and the equations for the
vertex, we get the integral equations
\begin{eqnarray}
    &&\left[E_{2}(\bm p)+E_{1}(\bm p)-W\right]\psi^+({\bm p})\nonumber\\&&=-\int\frac{d^3k}{(2\pi)^3}
    \left[V({\bm p},{\bm k},W)\psi^+(\bm k)
    +V({\bm p},{\bm k},W)\psi^-(\bm k)\right],\label{Eq: integral equations1}\ \ \ \ \\
         &&\left[E_{2}(\bm p)-E_{1}(\bm p)+W\right]\psi^-({\bm p})\nonumber\\&&=-\int\frac{d^3k}{(2\pi)^3}
    \left[V({\bm p},{\bm k},W)\psi^+(\bm k)
    +V({\bm p},{\bm k},W)\psi^-(\bm k)\right],\ \ \ \
    \label{Eq: integral equations}
\end{eqnarray}
where the potential is defined as
\begin{eqnarray}
    &&V({\bm p},{\bm
    k},W)\nonumber\\
    &&=-\frac{1}{\sqrt{2E_{1}(\bm p)~2E_{2}(\bm p)~2E_{1}(\bm k)~2E_{2}(\bm k)}}
    {\cal V}({\bm p},{\bm k},W).
\end{eqnarray}

The normalization of the wave function can be obtained by
the normalization of the vertex,
\begin{eqnarray}
    1=\int d^3p~\Gamma ^\dag({\bm p})
    \frac{\partial}{\partial W^2}\left[G({\bm p},W)\right]
    \Gamma ({\bm p})-R \label{la}
\end{eqnarray}
with
\begin{eqnarray}
    R&=&\int d^3p ~\int d^3p'\Gamma^\dag({\bm p})
        G({\bm p},W)
    \frac{\partial}{\partial W^2}
    [{\cal V}({\bm p},{\bm p}',W)]\nonumber\\&&\times
        G({\bm p}',W)
    \Gamma ({\bm p}').\ \
\end{eqnarray}

Usually, the one-boson-exchange model (OBE) is widely applied to study hadronic molecular states \cite{Liu:2008fh,Liu:2008tn,Liu:2007bf,Liu:2008xz,Shen:2010ky,Ding:2008gr,Ding:2009zq,Liu:2010hf}, which makes the above calculation simple; i.e., the potential is not dependent on
$W$, and $R=0$ \cite{Gross:1972ye}.

\subsection{Nonrelativized approximation of the CST}\label{t2}

In the following, we take the nonrelativized approximation in the CST, which was adopted in Refs~\cite{Gross:1969rv,Gross:1972ye,Buck:1979ff} to study the wave function of the deuteron. Using the nonrelativized approximation and the Fourier transform, the integral equations in Eqs. ~(\ref{Eq: integral equations1})-(\ref{Eq: integral equations}) can be transferred as

\begin{eqnarray}
    -\left(\frac{\nabla^2}{\mu}+\epsilon\right)\psi^+(\bm r)&=&
-\left[V(\bm r)+V(\bm r)F(\bm r)V(\bm r)\right]\psi^+(\bm r),\ \ \\
    \psi^-(\bm r)&=&-\left[F(\bm r)V(\bm r)\right]\psi^+(\bm
    r),\ \
\end{eqnarray}
where $F(\bm r)=[m_2-m_1+W+V(\bm r)]^{-1}$, and $E_{1}(\bm p)+E_{2}(\bm
p)-W\approx -\epsilon+\frac{ {\bm p}^2}{\mu}$ with the reduced mass
$\mu$ and the binding energy of two-body system $\epsilon=W-m_1-m_2$.

Assuming $m_{1,2}\gg \langle V\rangle$ as required from a loosely bound system, we obtain
\begin{eqnarray}
-\left(\frac{\nabla^2}{\mu}+\epsilon\right)\psi^+(\bm r)&=&-V(\bm r)\psi^+(\bm r),\label{hh1}\\
    \psi^-(\bm r)&=&0,
\end{eqnarray}
where Eq. (\ref{hh1}) corresponds to the Schr\"odinger equation.
The wave function in the momentum space
can be obtained by the Fourier transform, i.e.,
\begin{eqnarray}
\psi^+({\bm p})&=&\frac{1}{(2\pi)^{3/2}}\int d^3re^{-i{\bm p}\cdot{\bm
r}}\psi^+({\bm r}),
\end{eqnarray}
where ${\bm p}$ is the relative momentum.
The wave function $\psi^+({\bm p})$ satisfies
\begin{eqnarray}
    \int d^3p~|\psi^+(\bm p)|^2=1,
\end{eqnarray}
from the normalization condition of the vertex in Eq. (\ref{la}).

\subsection{Decay width}

With the preceding preparation, we illustrate how to calculate the decay width of the radiative and pionic open-charm decays. The general expression of the decay width is
\begin{eqnarray}
d\Gamma
&=&\frac{1}{2E}|{\cal M}|^2(2\pi)^4\delta^4\left(\sum_{i=1}^3q_i-P\right)\nonumber\\
&&\times
\frac{d^3q_1}{(2\pi)^32e_1}\frac{d^3q_2}{(2\pi)^32e_2}
\frac{d^3q_3}{(2\pi)^32e_3},
\end{eqnarray}
where $q_{i} (e_{i})$ ($i=1,2,3$) is the momentum (energy) of the final states. 

The decay amplitudes $\cal M$ for the sequential decay as shown in
Fig.~\ref{decay1} can be written as
\begin{eqnarray}
    {\cal M}&=&\frac{\mathcal{A}_{Y(4274)}~\mathcal{A}_{D_{s0}(2317)}}{m_{D_{s0}(2317)}^2-q^2},\label{Eq: amp}
\end{eqnarray}
where $\mathcal{A}_{Y(4274)}$ or $\mathcal{A}_{D_{s0}(2317)}$ denotes the vertex for the
decay of $Y(4274)$ or $D_{s0}(2317)$. $q$ and $m_{D_{s0}(2317)}$
are the four-momentum and mass of the intermediate state
$D_{s0}(2317)$. By the formulism in Section \ref{T1} and \ref{t2},
Eq. (\ref{Eq: amp}) can be further expressed in the center of mass
frame of the decaying particle $Y$ as
\begin{eqnarray}
\frac{\mathcal{A}_{Y(4274)}}{m_{D_{s0}(2317)}^2-q^2}
&=&\frac{\sqrt{2W(2\pi)^3}\sqrt{2E_{D_s}({\bm q})2E_{D_{s0}(2317)}({\bm q})}
}{W-E_{D_s}({\bm q})+E_{D_{s0}(2317)}({\bm q})}\psi^+({\bm q})\nonumber\\
&\approx&\sqrt{2M_{Y(4274)}}~~\sqrt{2(2\pi)^3\frac{m_{D_s}}
{m_{D_{s0}(2317)}}}~ ~\psi^+({\bm q}),\nonumber\\\label{11}
\end{eqnarray}
where $W$ is chosen as $M_{Y(4274)}$, which is the mass of
$Y(4274)$. The above expression is a bridge connecting the decay
amplitude and the wave function of the bound state. Thus, Eq. (\ref{11})
corresponds to the matrix element $\langle {D}_{s0}(2317)^-D^+_{s}
|\mathcal{H}_Y|Y(4274)\rangle$ in Eqs. (\ref{f1})-(\ref{f2}).

In addition, we also adopt the effective Lagrangians
\begin{eqnarray}
    &&\mathcal{L}_{D_{s0}(2317)D_{s}^{*}\gamma }=g_\gamma\,( \partial_\mu
    A_\nu\partial^\mu \phi_{D_s^*}^{\nu}-\partial_\nu
    A_\mu\partial^\mu \phi_{D_s^*}^{\nu} )\phi_{D_{s0}},\\
    &&\mathcal{L}_{D_{s0}(2317)D_{s}\pi }=g_\pi \phi_{D_s}\phi_{\pi}
    \phi_{D_{s0}}
\end{eqnarray}
to depict the interactions of $D_{s0}(2317)$ with $D^*_s\gamma$ and $D_s\pi$, respectively,
where $g_\gamma$ and $g_\pi$ are the effective coupling constants.
Thus, the amplitudes $\langle D_{s}^{*-}\gamma
|\mathcal{H}_\gamma|D_{s0}(2317)^-\rangle$ and $\langle D_{s}^-\pi^0|\mathcal{H}_\pi|D_{s0}(2317)^-\rangle$ read as
\begin{eqnarray}
&&\langle D_{s}^{*-}\gamma
|\mathcal{H}_\gamma|D_{s0}(2317)^-\rangle
=g_\gamma\,\varepsilon_{\gamma\mu}\varepsilon_{D^*_s\nu}
(g^{\mu\nu}q_\gamma\cdot q_{D_s^*}-q_\gamma^\nu q_{D_s^*}^\mu),\nonumber\\ \label{e1}\\
&&\langle D_{s}^\pm\pi^0|\mathcal{H}_\pi|D_{s0}(2317)^\pm\rangle
=g_\pi,\label{e2}
\end{eqnarray}
where $q_\gamma(q_{D_s^*})$ and
$\varepsilon_{\gamma}(\varepsilon_{D_s^*})$ are the momenta and the
polarization vectors of photon ($D^{*-}_s$), respectively. The
results in Eqs. (\ref{e1})-(\ref{e2}) correspond to
$\mathcal{A}_{D_{s0}(2317)}$ in Eq. (\ref{Eq: amp}).

\section{Numerical result}\label{sec3}

\begin{center}
\begin{figure}[htbp!]
\includegraphics[bb=50 45 610 275,scale=0.65,clip]{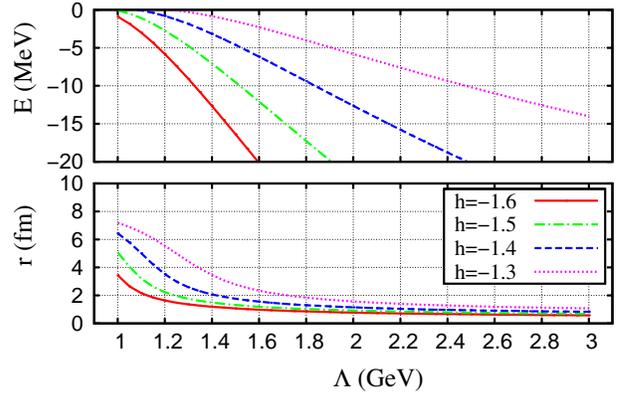}
\caption{(Color online.) The obtained bound state solutions (binding energy $E$ and root-mean-square $r$) of
$Y(4274)$ dependent on $h$ values and cutoff $\Lambda$.
\label{Fig: mass}}
\end{figure}
\end{center}

In Ref. \cite{Liu:2010hf}, the effective potential of $Y(4274)$ was obtained by the OBE model
\begin{eqnarray}
V(r)=\frac{2}{3}V_\eta^{Cross}(r)
&=&\frac{2}{3}\frac{h^2 q'^2_0}{f_\pi^2}Y(\Lambda,q'_0,m_\eta,r)
\label{potential}
\end{eqnarray}
with $Y$ function
\small
\begin{eqnarray}
Y(\Lambda,\kappa,m,r)=
 -\frac{1}{4\pi r}(e^{-\zeta_1r}-e^{-\zeta_2r})
+\frac{1}{8\pi}\frac{\zeta_2^2-\zeta_1^2}{\zeta_2}e^{-\zeta_2r}
\end{eqnarray}
\normalsize
and $\zeta_1=\sqrt{m^2-\kappa^2}$, $\zeta_2=\sqrt{\Lambda^2-\kappa^2}$,
$q'_0=m_{D_{s0}(2317)}-m_{D_s}$, $m_\eta=548$ MeV, and $f_\pi=132$~MeV, where
$Y(4274)$ is an S-wave $D_s\bar{D}_{s0}(2317)$ molecular charmonium.
By the OBE potential, one finds the bound state solutions for $Y(4274)$ shown in Fig.~\ref{Fig: mass}, where
the dependence of the binding energy and the root-mean-square radius on different values of $h$ and $\Lambda$ are given (see Ref. \cite{Liu:2010hf} for more details).

\begin{figure}[htbp!]
\includegraphics[bb=65 5 410 280,scale=0.75,clip]{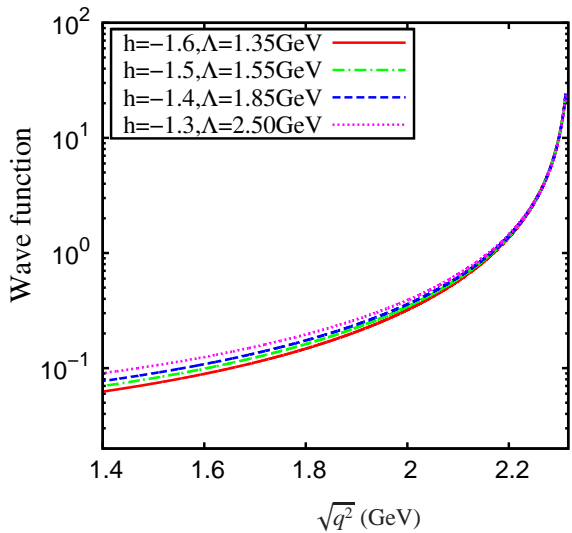}\put(-140,5){ $\sqrt{q^2}$~(GeV)}
\caption{(Color online.) The dependence of $\psi^+({\bm q})$ on
$\sqrt{{q}^2}$. Here, $\sqrt{q^2}$ corresponds to the invariant mass $M_{D_s^*\gamma}$ or
$M_{D_s\pi}$ as shown in Fig.~\ref{decay1}. \label{Fig: wf}}
\end{figure}

For the decay of $Y(4274)$ as shown in Fig.~\ref{decay1}, in the center of mass frame of $Y(4274)$, there exists a
relation between the four-momentum and three-momentum of the off-shell
intermediate state $D_{s0}(2317)$, i.e.,
 ${q}^2=M_{Y(4274)}^2+m_{D_s}^2-2M_{Y(4274)}\sqrt{m_{D_s}^2+{\bm
q}^2}$, where $\sqrt{{q}^2}$ also corresponds to the
invariant mass $M_{D_s^*\gamma}$ or $M_{D_s\pi}$. In Fig.
\ref{Fig: wf}, we list the variation of wave function $\psi^+(\bm
q)$ obtained by the OBE model with $\sqrt{{q}^2}$ when taking different $h$
values, where we can reproduce the binding energy
($M_Y-m_{D_{s0}(2317)}-m_{D_s}\approx-11$ MeV) with the corresponding $\Lambda$
values.
 The results in
Fig.~\ref{Fig: wf} indicate that $\psi^+({\bm q})$ is not strongly
dependent on $h$ and $\Lambda$.

Besides information on the wave function of the
$D_{s}\bar{D}_{s0}(2317)$ molecular state, the coupling constants
$g_{\gamma}$ and $g_\pi$ can be extracted by the relations
\begin{eqnarray}
\Gamma\left(D_{s0}^{-}(2317)\rightarrow D_{s}^{*-}\gamma\right)
&=&g_\gamma^2 \frac{|{\bm q_{D_s^*}}|}{4\pi M_{D_{s0}(2317)}^2}~(q_{D_s^*}\cdot q_\gamma)^2,\\
\Gamma\left(D_{s0}^{\pm}(2317)\rightarrow D_{s}^\pm\pi^0\right)
&=&g_\pi^2\frac{|{\bm q}_{D_s}|}{8\pi M_{D_{s0}(2317)}^2},
\end{eqnarray}
where we use theoretical values to determine the constants, since these experimental partial decay widths are absent at present.
In the literature \cite{Faessler:2007gv,Lutz:2007sk,Guo:2008gp,Colangelo:2003vg,Godfrey:2003kg,:2003dpa,Bardeen:2003kt,Lu:2006ry,Wei:2005ag,Close:2005se,Liu:2006jx,Wang:2006mf,Colangelo:2005hv,Nielsen:2005zr,Cheng:2003kg,Azimov:2004xk},
the radiative and pionic decays of $D_{s0}(2317)$ were calculated under different structure assignments to $D_{s0}(2317)$.
In this work, we take the typical values $\Gamma\left(D_{s0}(2317)^-\to D_s^{*-}\pi\right)$ =10~keV \cite{Godfrey:2003kg} and $\Gamma\left(D_{s0}(2317)^-\to
D^{*-}\gamma\right)$ =1~keV \cite{Close:2005se}, where $D_{s0}(2317)$ is assumed to be a charm-strange meson.
{We emphasize that our numerical result is dependent on the assumption on the partial decay width of $D_{s0}(2317)$, since different structure assignments to $D_{s0}(2317)$ can result in different partial decay widths of $D_{s0}(2317)$. Discussing the structure of $D_{s0}(2317)$ is beyond the scope of this work. If the partial decay widths of $D_{s0}(2317)$ are determined in the future, the corresponding decay widths of $Y(4274)$ open-charm radiative and pionic decays can be obtained by the present results multiplied by an extra factor.}

\begin{table}[h!]
    \caption{The radiative and pionic open-charm decay widths of $Y(4274)$ with typical values of $h$ and $\Lambda$.\label{Tab: decay width}}
\small
\renewcommand\tabcolsep{0.4cm}
\begin{center}
    \begin{tabular}{ c | cccc}  \toprule[1pt]
    $h$& $\Lambda$~(GeV) & $\Gamma_\gamma$~(keV) &
    $\Gamma_\pi$~(keV)\\\midrule[1pt]
   -1.60  &  1.35 & 0.049 & 0.74 \\
   -1.50  &  1.55 & 0.050 & 0.78 \\
   -1.40  &  1.85 & 0.050 & 0.75 \\
   -1.30  &  2.50 & 0.049 & 0.77 \\\bottomrule[1pt]
\end{tabular}
\end{center}
\end{table}

In Table~\ref{Tab: decay width}, we give the radiative and pionic
open-charm decay widths for $Y(4274)$ with different combinations of $h$ and $\Lambda$ values.
We find that these obtained partial decay widths of $Y(4274)$ are
not sensitive to the corresponding combination of $h$ and $\Lambda$.
The calculated decay widths of $Y(4274)\to
D^{+}_{s}D^{*-}_{s}\gamma$ and $Y(4274)\to D^+_{s}D^-_{s}\pi^0$ are
around 0.05 keV and 0.75 keV, respectively. Of course, we admit
that the above results are dependent on the decay widths of
$D_{s0}^{-}(2317)\rightarrow D_{s}^{*-}\gamma$ and
$D_{s0}^{-}(2317)\rightarrow D_{s}^-\pi^0$, since the coupling
constants $g_\gamma$ and $g_\pi$ are determined by these theoretical
values. We emphasize that in the current work the
dissociation of $Y(4274)$ is described by the wave function
$\psi^+({\bm q})$. Thus, estimating the branching ratios of
$Y(4274)$ decays under the CST is a realistic approach, since the
wave function adopted in the calculation is obtained by solving the
Schr\"{o}dinger equation with the OBE potential in Eq.
(\ref{potential}).

\begin{figure*}[hbtp!]
\begin{center}
\begin{tabular}{cc}
\includegraphics[bb=60 5 410
280,scale=0.6,clip]{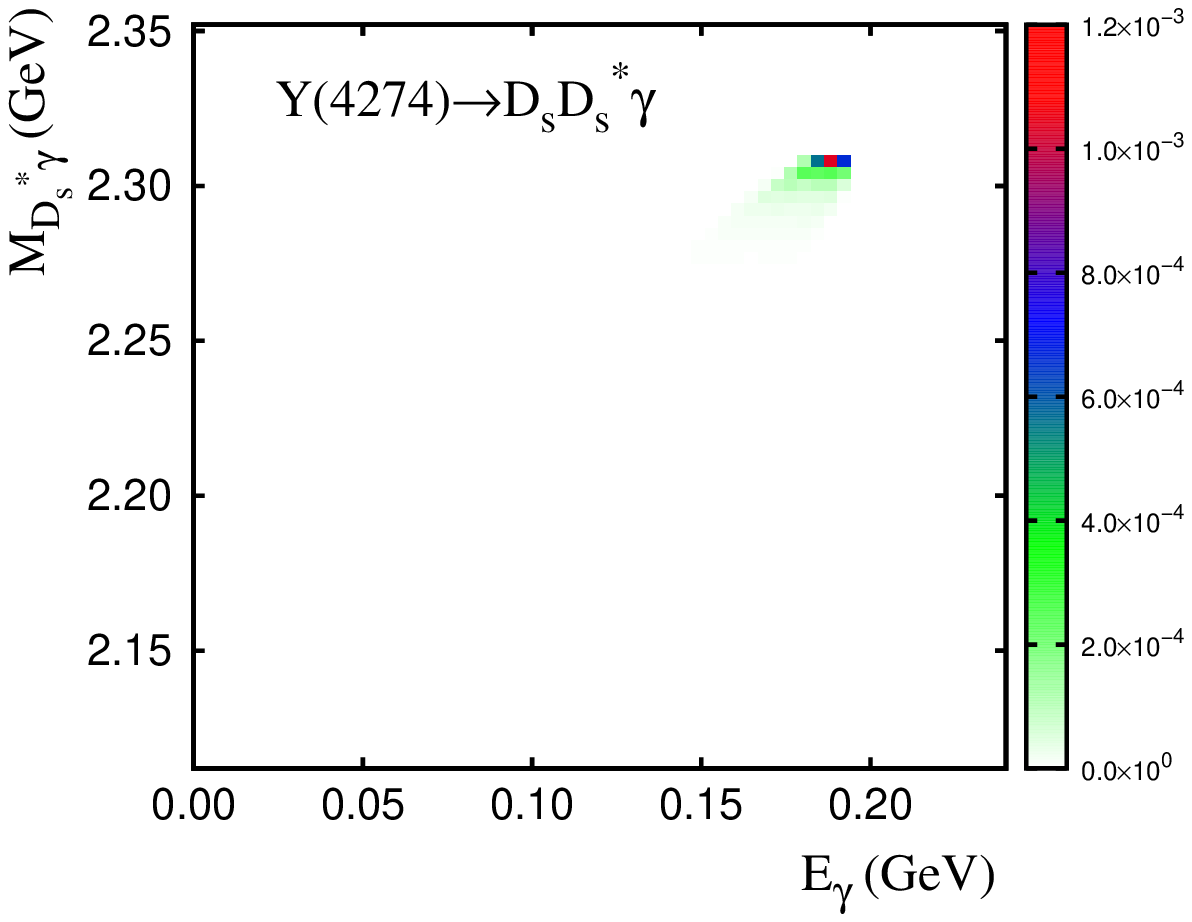}&\includegraphics[bb=60 5 410 280,scale=0.6,clip]{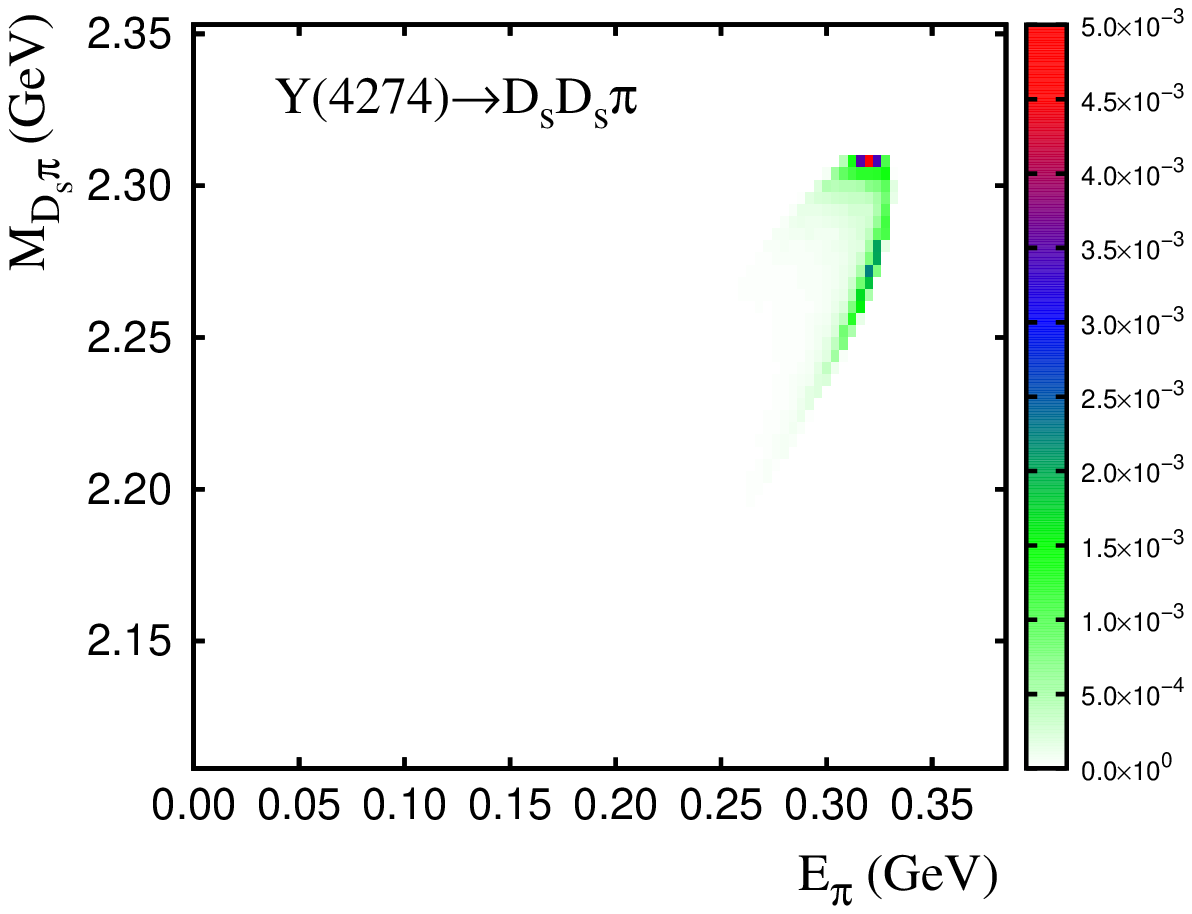}\\
\includegraphics[bb=60 5 410
280,scale=0.6,clip]{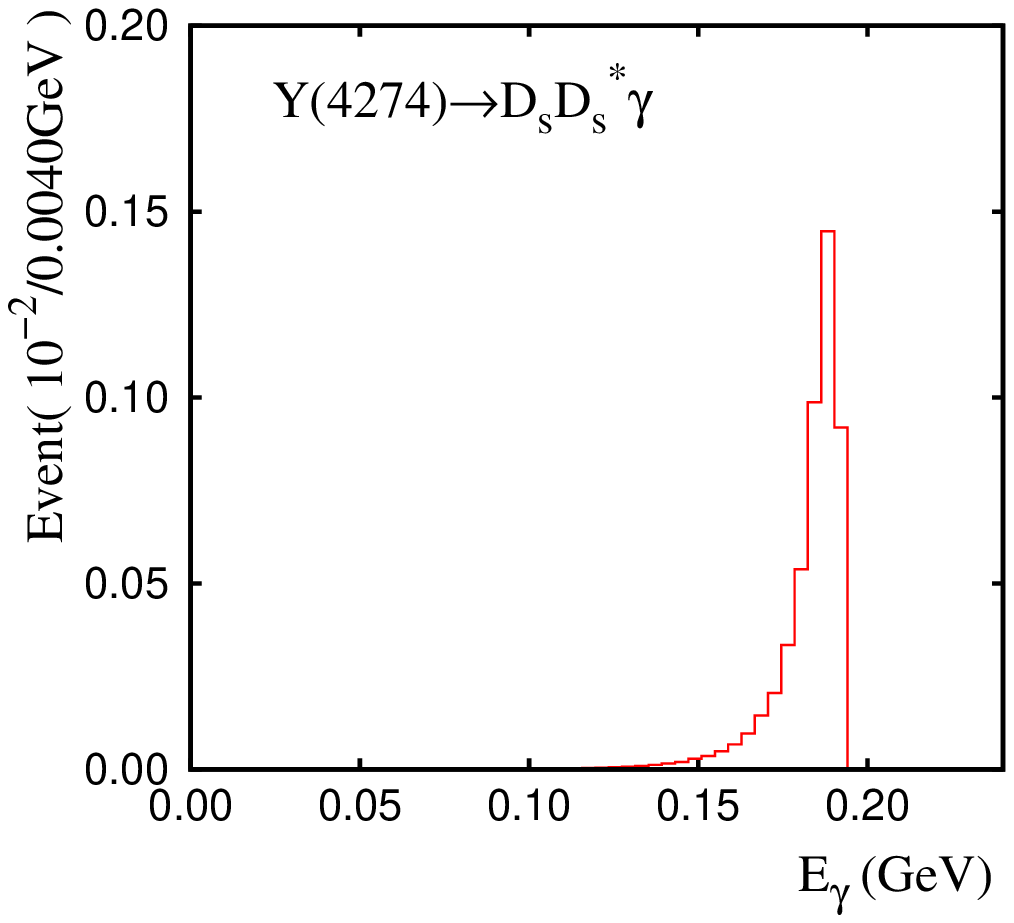}&\includegraphics[bb=60 5 410 280,scale=0.6,clip]{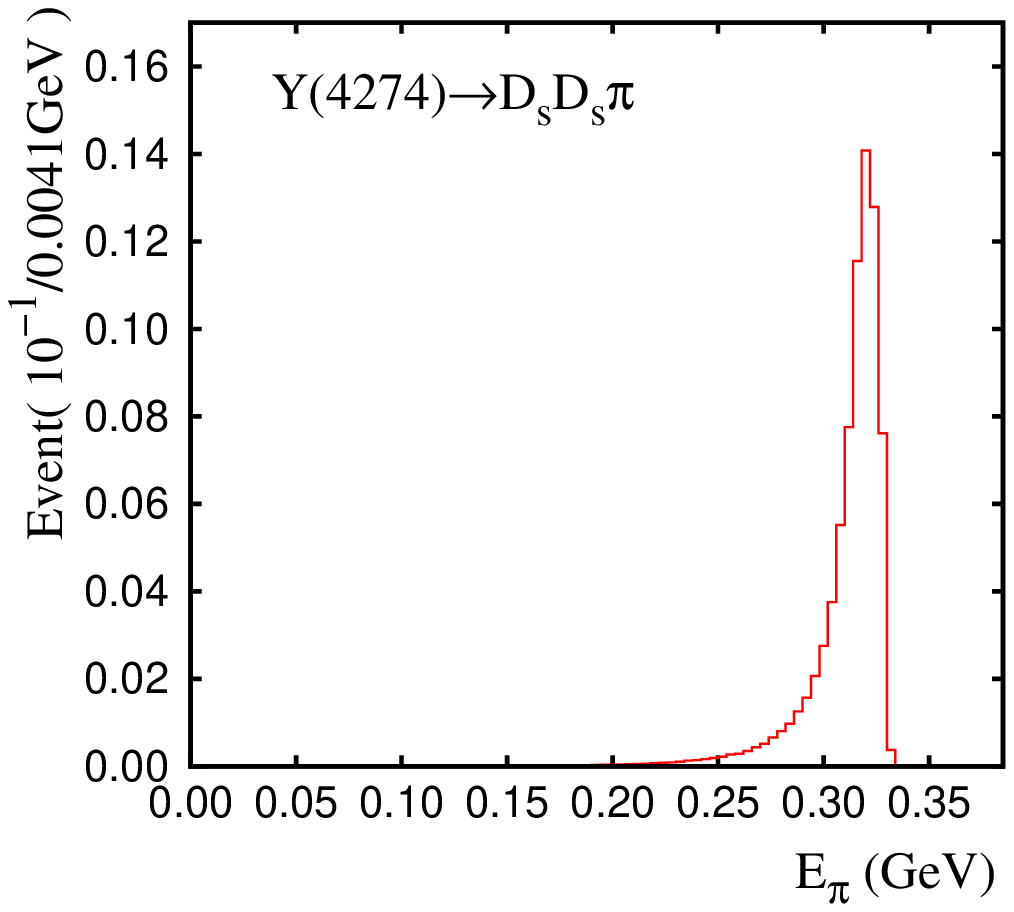}
\end{tabular}
\end{center}
\caption{(Color online.) The Dalitz plot analysis and the photon and
pion spectra for $Y(4274)\to{D}^{+}_s D^{*-}_s\gamma$ and
$Y(4274)\to{D}^{+}_s D_s^-\pi^0$ decays. Here $h=-1.3$ and
$\Lambda=2.5$~GeV. \label{decay}}
\end{figure*}

{In fact, we expect that the decays considered in this paper would be insensitive to details of the wave function. The reason is that the binding energy -11 MeV is much smaller than the range of forces, which is given by the mass of the eta in our model. This means that the size of the molecule is around 1/(160 MeV). The decay channels through the decays of one constituent should "know" little about the short-distance behavior of the wave function, since they are dominated by the long-distance part. This fact can be seen easily from Table \ref{Tab: decay width}, where different sets of parameters give almost the same results for the decay widths. One may further check Fig. \ref{Fig: wf}; the long-distance part of the wave function is the same for different parameters, while one sees clear differences at short distances, which correspond to small values of $q^2$ and hence large values of ${q}^2$. The observation of similar features shown in Fig. \ref{decay} should be strong evidence of an intermediate state with a well-defined mass of $~2.3$ GeV.}

Besides presenting the decay widths of $Y(4274)$, we also carry out the study of the line shapes of the
photon and pion spectra of $Y(4274)\to{D}^{+}_s D^{*-}_s\gamma$ and
$Y(4274)\to{D}^{+}_s D_s^-\pi^0$ processes under the assignment of the
$D_s\bar{D}_{s0}(2317)$ molecular state to $Y(4274)$. Here, we use the CERNLIB
program FOWL to produce the Dalitz plots and the line shapes of the photon and pion spectra
of $Y(4274)\to{D}^{+}_s D^{*-}_s\gamma$ and
$Y(4274)\to{D}^{+}_s D_s^-\pi^0$ (see Fig. \ref{decay}).

For $Y(4274)\to D_s^+\bar{D}_s^*\gamma$, an accumulation appears in
the Dalitz plot with photon energy $E_\gamma\sim 0.19$ GeV and
$M_{D_s^*\gamma}\sim 2.3$ GeV, {which also reflects that $D_s^*\gamma$ and $D_{s}\pi$ are from an intermediate state $D_{s0}(2317)$.}
The line shape of the photon
spectrum of $Y(4274)\to D_s^+\bar{D}_s^*\gamma$ indicates that
a very sharp peak exists near the large end point of photon energy,
which directly corresponds to the accumulation in the Dalitz plot.
The result of the line shape of the pion spectrum for the 
$Y(4274)\rightarrow {D}_{s}^+ D_{s}^-\pi^0$ process shows that a
steep peak also exists near the large end point of pion energy, which
is similar to the situation of $Y(4274)\to D_s^+{D}_s^{*-}\gamma$.


\section{Discussion and conclusion}\label{Sec: Sum}

More and more observations of charmonium-like states $X$, $Y$, $Z$ in $B$ meson decays are providing us with a good platform to intensively study the internal mechanism for producing these charmonium-like states. this is one of the most important research topics-full of challenges and opportunities-in charm physics \cite{Li:2008ey}.

Stimulated by the recent evidence of $Y(4274)$ \cite{Collaboration:2011at,Aaij:2012pz} and the proposed $D_s\bar{D}_{s0}(2317)$ molecular charmonium explanation for $Y(4274)$ \cite{Liu:2010hf}, in this work we study the open-charm radiative and pionic decays $Y(4274)\to D_s^+{D}_s^{*-}\gamma$ and $Y(4274)\to D_s^+D_s^-\pi^0$, where we not only present the calculation of its decays but also give the line shape of the photon and pion spectra of
$Y(4274)\to D_s^+{D}_s^{*-}\gamma$ and $Y(4274)\to D_s^+D_s^-\pi^0$ in detail. These theoretical predictions of the decay behavior of $Y(4274)$ provide the information for further experimental studies of $Y(4274)$.

We emphasize that $Y(4274)\to D_s^+D_s^-\pi^0$ is peculiar to $Y(4274)$ under the assignment of $D_s\bar{D}_{s0}(2317)$ molecular state, since this decay reflects the internal structure $Y(4274)$.
$Y(4274)\to D_s^+D_s^-\pi^0$ occurs via the interaction of $D_{s0}(2317)^-$ with $D_s^-\pi^0$ after the collapse of  $Y(4274)$ into $D_s^+$ and $D_{s0}(2317)^-$, where $D_{s0}(2317)^-\to D_s^-\pi^0$ is a special decay channel observed by experiment \cite{Aubert:2003fg}. Thus, the experimental search for a  $Y(4274)\to D_s^+D_s^-\pi^0$ channel and the measurement of the line shape of the pion spectrum of $Y(4274)\to D_s^+D_s^-\pi^0$ will be an interesting research topic.

We must admit that the predicted partial decay widths of $Y(4274)$ are tiny compared with its total width, which makes the branching ratios of the considered decays about $10^{-6}\sim 10^{-5}$, since the channels considered in this paper are either electromagnetic or isospin violation processes. Although these decay channels of $Y(4274)$ cannot be the ideal ones according to the present status of CDF and LHCb, searching for open-charm radiative and pionic decays of $Y(4274)$ could be considered topics for possible efforts in future experiments, such as the forthcoming BelleII \cite{belle} and SuperB \cite{superb}.

\section*{Acknowledgement}

We would like to thank the referee for his/her useful suggestion on the discussion about the numerical result.
This project is supported by the National Natural Science Foundation
of China under Grant Nos. 111750731, 10905077, and 11035006, the Ministry of Education of China (FANEDD under Grant No.
200924, DPFIHE under Grant No.  20090211120029, NCET under Grant No.
NCET-10-0442, the Fundamental Research Funds for the Central
Universities, the Fok Ying-Tong Education Foundation (No. 131006), the project sponsored by SRF for ROCS, SEM under Grant
No. HGJO90402), and the Chinese Academy of Sciences (the Special Foundation of the
President under Grant No. YZ080425).

\end{document}